\documentclass[amsmath,twocolumn,superscriptaddress,prl,aps]{revtex4}
\usepackage{amsthm,amsfonts,graphicx,verbatim, color}
\usepackage{graphicx}
\usepackage{bm}
\usepackage[utf8]{inputenc}
\let\oldmarginpar\marginpar
\renewcommand\marginpar[1]{\-\oldmarginpar[\raggedleft\footnotesize #1]%
{\raggedright\footnotesize #1}}

\newcommand{\be}{\begin{equation}}
\newcommand{\ee}{\end{equation}}
\newcommand{\bea}{\begin{eqnarray}}
\newcommand{\eea}{\end{eqnarray}}

\renewcommand{\epsilon}{\varepsilon}
\renewcommand{\vec}[1]{{\bf #1}}

\renewcommand{\cite}[1]{[\onlinecite{#1}]}

\newcommand{\sign}{\mathrm{sign}}

\begin{document}

\title{Dirty Weyl Fermions: rare region effects near 3D Dirac points}
\author{Rahul Nandkishore}
 \affiliation{Princeton Center for Theoretical Science, Princeton University, Princeton, New Jersey 08544, USA}
\author{David A. Huse}
 \affiliation{Princeton Center for Theoretical Science, Princeton University, Princeton, New Jersey 08544, USA}
 \affiliation{Department of Physics, Princeton University, Princeton New Jersey 08544, USA}
\author{S. L. Sondhi}
 \affiliation{Department of Physics, Princeton University, Princeton New Jersey 08544, USA}

\begin{abstract}
We study three-dimensional Dirac fermions with weak finite-range scalar potential disorder. We show that even though disorder is perturbatively irrelevant at 3D Dirac points, nonperturbative effects from rare regions give rise to a nonzero density of states and a finite mean free path,
with the transport at the Dirac point being dominated by hopping between rare regions. As one moves in chemical potential away from the Dirac point, there are interesting intermediate-energy regimes
where the rare regions produce scattering resonances that determine 
the DC conductivity.
We also discuss the interplay of disorder with interactions at the Dirac point.  
Attractive interactions drive a transition into a granular superconductor, with a critical temperature that depends strongly on the disorder distribution.  In the presence of Coulomb repulsion and weak retarded attraction, the system can be a Cooper-pair glass. Our results apply to all 3D systems with Dirac points, including Weyl semimetals, and overturn a thirty year old consensus regarding the irrelevance of weak disorder at 3D Dirac points. 
\end{abstract}
\maketitle

The discovery of two-dimensional (2D) Dirac systems such as graphene and the surface states of topological insulators has sparked an explosion of activity in condensed matter physics \cite{CastroRMP, KaneHasan}. Such materials, which are gapped everywhere except at isolated points in the Brillouin zone, play host to an abundance of new physics. In particular, when the chemical potential is placed at the `Dirac point', they display behavior that is intermediate between metals and insulators, in that the spectrum is gapless, but displays a vanishing low-energy density of states (DOS). The recent proposal \cite{Murakami, Wan, Balents, Turner}  (and potential discovery \cite{Turner}) of Weyl semimetals provides a three-dimensional (3D) version of this behavior, and promises to ignite a blaze of interest in 3D Dirac points.

The consensus in the theory literature, from original work by Fradkin \cite{Fradkin} in the 1980's to more recent work on Weyl semimetals \cite{Goswami, Hosur, Garate}, is that weak disorder is perturbatively irrelevant at 3D Dirac points, implying that sufficiently weak disorder does not affect the vanishing of the density of states (DOS) at the Dirac point, or the divergence of the mean free path. 
However, all existing theory works on this subject have ignored non-perturbative rare region effects, which can dominate the 
physics at particle-hole symmetric points \cite{DiracSC, Lesik}. A study of disordered 3D Dirac points that incorporates rare region effects is thus an interesting and timely task.

In this Letter we show, in a direct contradiction of the prevailing theory consensus, that the disordered system has a non-vanishing DOS and a finite mean free path at the Dirac point for arbitrarily weak disorder. Our results differ from previous calculations because we take into account non-perturbative rare region effects,
which have been neglected in all previous (disorder-averaged) calculations.  
We discuss the multiple distinct transport regimes that arise as the doping away from the Dirac point is varied. We also discuss the influence of interactions, following our earlier analysis of the
dirty $2D$ Dirac semimetal \cite{DiracSC}.  We show that 
weak attractive interactions can drive a transition to a (granular) superconductor, with a non-universal critical temperature. In the presence of Coulomb repulsion and retarded attractive interactions, the system can be a Cooper pair glass, with infinite superconducting susceptibility but no long range phase order. Our calculation is done for a Weyl semimetal, but the results apply to all 3D Dirac points.

{\it The model:}
The low energy Hamiltonian of the clean, non-interacting Weyl semimetal is
\begin{equation}
H = \sum_{a = 1}^{2N} \sum_{i=1}^{3} v^a_i \psi^{\dag}_{a}  \sigma_i  k_i \psi_{a} ~,
\end{equation}
where the two-component spinor $\psi_a(\vec{k})$ represents a state near the Dirac node $a$, with a momentum $\vec{k}$ relative to the Dirac point. Dirac nodes always come in symmetry related pairs. For simplicity we consider the minimal model \cite{Balents}, which has only two Dirac nodes at momenta $\pm \vec{Q}$, although the analysis can be easily generalized without altering the essential results. The Dirac points are topologically protected in the absence of inter-node scattering. In general the dispersion about the Dirac nodes is anisotropic, but for simplicity we consider the isotropic limit $v_1=v_2=v_3 = v$. The DOS (per unit volume) at low energies vanishes as $\nu(E) = \frac{N E^2}{2 \pi^2 (\hbar v)^3}$, where $N$ is the number of Dirac points.

We now consider adding weak quenched scalar potential disorder to the system (strong disorder has been studied in \cite{Arovas}). The perturbative effect of weak disorder on the electron Green function can be determined \cite{Abrikosov-Gorkov, Hosur} by evaluating the electron self energy $\Sigma$, which yields
$ \Sigma(\omega, \vec{k}\rightarrow 0) \sim V \omega^2$, where $V$ is the mean square scalar potential. This vanishes more rapidly than $\omega$ at low energies and allows existence of sharp quasiparticles. Similarly, a self consistent Born approximation (SCBA) for the mean free path $l$ leads to
\begin{equation}
\frac{\hbar v}{l} = \frac{\hbar v}{l} V \int_0^{\Lambda} \frac{\nu(E) dE}{E^2 + \hbar^2v^2/l^2}  \label{eq: SCBA}
\end{equation}
when the chemical potential is at the Dirac point; $\Lambda$ is a UV cutoff and $\nu(E) \sim E^2$. For sufficiently weak disorder $V \rightarrow 0$, this admits only the trivial solution $1/l = 0$ (at small non-zero $\mu$, $l$ diverges as $l \sim 1/(V\mu^2)$ within SCBA). This is in sharp contrast to two-dimensional Dirac materials, where, within SCBA, disorder produces a crossover to diffusive behavior at long length scales \cite{Ando}. The difference arises because the DOS vanishes more rapidly in $3D$, making disorder perturbatively irrelevant instead of marginal \cite{internode}.

{\it Density of states from rare regions:} We now show that the low-energy DOS of dirty Dirac fermions is non-zero, because of resonances arising on rare regions. We first present a heuristic argument that captures the basic results, before providing a more detailed analysis. Since rare region effects can be sensitive to the distribution of disorder, we consider two distinct models of disorder. Model A consists of unbounded disorder - the chemical potential is correlated over a length scale $R$ (of order the lattice constant), with a Gaussian distribution for the local scalar potential $P(U) \sim \exp(- U^2 / 2 \mu_0^2)$. Meanwhile, Model B is a model with bounded disorder - we have regions of linear size $R$, with local chemical potential randomly $U= \pm \mu_0$. The mean square `average disorder strength' in either model is $V = \mu_0^2 R^3$. While `real' disorder is more complex than these toy models, we believe the basic results, (e.g. non-zero DOS at the Dirac point), are generic.

First, we consider Model A. A straightforward application of the central limit theorem implies that the probability
that a given compact region of volume $L^3$ has an average potential $U$, $P_A(U, L)$, is given by
\begin{equation}
P_A(U, L) = \frac{L^{3/2}}{R^{3/2} \mu_0 \sqrt{2 \pi}} \exp \left( - L^3 U^2 / 2 R^3 \mu_0^2\right) ~, \label{eq: PA}
\end{equation}
where $\mu_0$ is the fundamental parameter controlling the disorder strength. 
%
Such a rare region, if large enough, has a local Dirac point that is shifted by an amount $U$.
The amount $U$ defines a wavelength $2 \pi \hbar v/U$.
If $L$ is larger than this length, then such regions contribute on average $\sim U^2 L^3$ to the DOS at $E=0$, because of their local scalar potential.
The total DOS at zero energy, $\nu_0$, is obtained by integrating over all such regions:
%
\begin{equation}
\nu_0 \sim  \int_R^{\infty} dL \int_{2\pi \hbar v / L}^{\infty} dU  U^2 L^3 P_A(U, L) ~, 
\end{equation}
where $P_A$ is given by (\ref{eq: PA}). In a saddle point approximation, this integral is dominated by its lower limits
$U \sim 2\pi \hbar v/L$ and $L \sim R$, and the zero energy DOS is
\begin{equation}
\nu_0 \sim \exp \left(-  A (\hbar v/\mu_0 R)^2\right) ~,
\label{eq: dosunbounded}
\end{equation}
which is exponentially small in weak disorder \cite{footnote}. 
This is dominated by the smallest regions of order the cutoff $R$, and we expect that $A$ is a nonuniversal constant of order one set by details near the cutoff.
In the formal white noise limit ($R \rightarrow 0$ at constant $V = \mu_0^2 R^3$), 
these regions are not exponentially rare  \cite{bounded}, however, in any real material the formal white noise limit is unattainable, and $R$ has a minimum size of order $\hbar v /\Lambda \approx a$, where $\Lambda$ is the bandwidth and $a$ is the lattice spacing.

We now repeat the above calculation with bounded disorder (Model B). Now the DOS arises from large regions of size $L \ge 2\pi\hbar v/\mu_0$ with nearly uniform scalar potential,
\begin{equation}
\nu_0 \sim \int_{2\pi\hbar v/\mu_0}^{\infty} dL \mu_0^2 L^3 2^{- L^3/R^3} \sim  \exp \left(-A' (\frac{\hbar v}{R\mu_0})^3\right) \label{eq: dosbounded} ~,
\end{equation}
where again $A'$ is an undetermined numerical constant, and the integral has been evaluated in a saddle point approximation and is dominated by the lower limit. This is parametrically smaller than
the corresponding DOS with unbounded disorder (\ref{eq: dosunbounded}). We note too that the estimate (\ref{eq: dosbounded}) takes into account only rare regions that are close to spherical in shape. We neglect irregular shaped regions because we expect that they will need to be larger to support bound states, and will thus be exponentially rarer. Nevertheless, the contribution of irregular shaped regions is an interesting topic for future work.


{\it Transport in the lower energy regimes:} On energy scales $|E| > \sqrt{(\hbar v)^3\nu_0}$, the dominant contribution to the DOS comes from the extended states. 
Carriers with energies closer to the Dirac point spend most of their time in resonant states on the rare regions that have DOS $\nu_0$.
The hopping $t(r)$ between such rare regions a distance $r$ apart can be extracted from the Green function $G(t,\vec{r})$ for extended states according to
\begin{eqnarray}
t(r) &\sim& \int_{-\infty}^{\infty} dt G(t,r) 
= \int d^d k G(0, \vec{k}) e^{-i \vec{k} \cdot \vec{r}} \\ &=& \int_0^{\Lambda} k^2 dk \int_{-1}^{1} d \cos \theta \frac{e^{-i k r \cos \theta}}{v k } 
= \frac{i}{v r^2} (1 - \cos \Lambda r) \sim 1/r^2\nonumber
\end{eqnarray}
where $\Lambda$ is a UV cutoff and we assume the energy is at the Dirac point. Thus, the hopping amplitude between rare regions falls off as $t(r)\sim 1/r^2$.
Hopping will be effective between two rare regions with energy difference less than the hopping amplitude.  For any one such region hopping will occur, since at distance $r$ there is typically another region with energy within $\sim 1/(r^3\nu_0)$, which falls off faster than $t(r)$.  Thus we can conclude that these low energy carriers near the Dirac point are not strongly Anderson localized, in spite of hopping among randomly-placed rare regions of random energies.

%
\begin{figure}
\includegraphics[width = 0.9 \columnwidth]{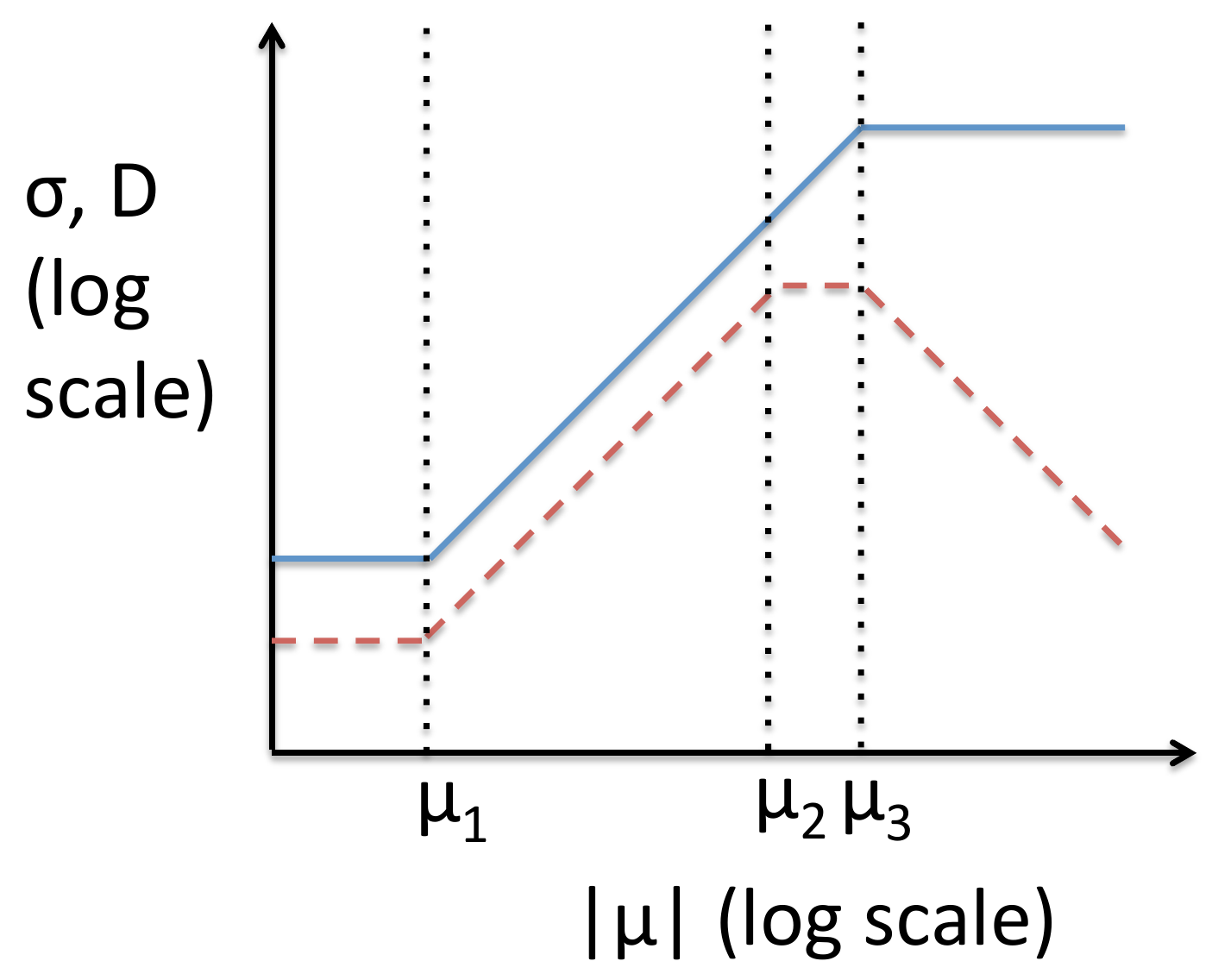}
\caption{\label{fig: conductivity} Schematic behavior of the zero-temperature DC conductivity $\sigma$ (solid blue line) and diffusivity $D$ (dashed red line)
as a function of the chemical potential $\mu$ for disordered non-interacting massless 3D Dirac fermions.  The dotted vertical lines are guides to the eye. Moving from low to high energy, the sequence of regimes and their boundaries is: hopping regime, $\mu_1\sim(\hbar v)^2\nu_0 b$, intermediate regime I, $\mu_2\sim(\hbar v)^{3/2}\nu_0^{1/2}$, intermediate regime II,
$\mu_3\sim(\hbar v)^{5/2}(\nu_0 b/V)^{1/2}$, SCBA regime.  The density of states is $\nu_0$ in the first two regimes, where it is dominated by rare regions of
linear size $b$.  $V$ is the mean square random scalar potential.  The rare regions dominate the scattering for all regimes other than the highest-energy SCBA regime.
The nonzero slopes on this log-log plot are $\pm 2$.
For more details, see text and \cite{supplement}.}
\end{figure}

{\it Resonant rare regions from Dirac equation}: We can obtain more quantitative results
by directly solving the 3D massless Dirac equation (1) in a scalar potential \cite{supplement}. 
We consider a spherical volume of radius $\rho\rightarrow\infty$, with a scalar potential $V(r) = \hbar v\lambda$ for $0 < r < b$, and $V(r) = 0$ for $r > b$. 
At energy $E=0$ there are particular values of $\lambda = \lambda^{(n)}_c \approx n \pi / b $ that give rise to bound states (here $n$ is any non-zero integer).  For $r>b$ these bound-state wavefunctions do indeed fall off as $\sim 1/r^2$.  Moving away from $E=0$ these states are now within a continuum of extended states, so they become resonances: bound states with a finite lifetime, with the lifetime diverging as $\sim\frac{\hbar^2 v}{E^2 b}$ as $E\rightarrow 0$.  At weak disorder the low energy DOS $\nu_0$ is dominated by the resonances with $n=\pm 1$, which are exponentially more probable than the others.


Meanwhile, solving the scattering problem at low energies, 
we find a cross section that scales as
\begin{equation}
\sigma(E, \lambda) \sim \frac{\big(\frac{E}{\hbar v}\big)^2 b^2}{(\lambda - \lambda_c^{(n)}(E))^2 + \big(\frac{E}{\hbar v}\big)^4 b^2} ~,
\end{equation}
where $\lambda_c^{(n)}(E) - \lambda_c(0) \sim \frac{E}{(\hbar v)}$. Thus, there are lines of resonances in the $\lambda, E$ plane, with width $\delta \lambda \sim \delta E \sim b E^2$. 

Finally, extracting the mean free path from the scattering cross section, we conclude that there are four distinct regimes of transport for these noninteracting carriers \cite{supplement}.
For $|E| > (\hbar v)^{5/2}\sqrt{\nu_0b/V}$, transport and the DOS are dominated by extended states. The DOS scales as $\nu \sim E^2$ and the mean free path scales
as $l \sim 1/E^2$. Meanwhile $|E| < (\hbar v)^2\nu_0 b$ 
is the hopping regime, discussed above, where the transport is via hopping between bound states on rare regions (with typical hopping rate $\sim\nu_0^2$),
and the typical hop is over a distance $\sim \nu_0^{-1}$. 
There are two intermediate-energy regimes.  In the intermediate regime with lower $|E|$,
both the DOS and scattering are still dominated by rare regions. Transport proceeds via `extended' states that get absorbed on a resonant rare region,
stay on the rare region for a time $\sim E^{-2}$, then get re-emitted, moving a distance $l \sim \nu_0^{-1}$ before being re-absorbed and re-emitted.
In this intermediate energy regime, even though transport proceeds through `extended states' the carriers spend most of their time trapped on the rare resonances.
Meanwhile, in the other intermediate regime of higher $|E|$, 
the DOS is dominated by extended states, but the mean free path $l$ is still limited by resonant scattering events, even though electrons now spend only a small fraction of their time trapped on the resonances. Applying standard techniques such as continuous time random walks and the Einstein-Smoluchowski relation \cite{supplement}, we predict that the conductivity and diffusion co-efficient for the disordered system should behave as shown in Fig. \ref{fig: conductivity}.


This concludes our discussion of rare region effects about \emph{non-interacting} 3D Dirac points. We now turn to the effect of interactions.

{\it Repulsive interactions:} Above a critical interaction strength, repulsive interactions destroy the Weyl semimetal phase \cite{Wei, Zhang}. Subcritical repulsive interactions suppress (charged) rare regions, and reduce the rare-region DOS at the Dirac point. We defer further consideration of repulsive interactions to future work. 

{\it Attractive interactions}
Attractive interactions above a critical strength will trigger superconductivity in the clean system \cite{Meng, Cho}. Subcritical interactions will produce
local pairing on rare regions where the local DOS is non-zero over a larger length scale than the local coherence length $\xi$. Establishment of phase coherence between
islands by Josephson coupling will then drive the system into a (granular) superconducting state at sufficiently low temperatures. We have discussed similar phenomena for the 2D Dirac semimetal in \cite{DiracSC}. 
We focus on estimating the energy scale for the superconducting state, first for Model A disorder and then for model B.

Local pairing occurs in islands of local doping $\mu$ and size $L \ge \xi$, where $\xi \sim (v/\omega_D) \exp(\hbar^3 v^3/ g \mu^2)$
is the local coherence length in the BCS approximation, $\omega_D$ is the Debye frequency and $g$ is the strength of the attraction in the leading pairing channel. 
Integrating over $L$ in a saddle point approximation, we find the result is dominated by islands of size $L \cong \xi$ (assuming $R\ll\xi$).
For model A disorder, the probability of finding such an island is
\begin{equation}
P^A_{SC} \sim  \int_0^{\min(\Lambda,\frac{(\hbar v)^{3/2}}{g^{1/2}})} d \mu \exp\left(- \frac{ \mu^2 v^3}{2\mu_0^2 \omega_D^3 R^3} \exp (3\hbar^3 v^3 / g \mu^2) \right) ~,
\end{equation}
where $\Lambda$ is the bandwidth, and $ (\hbar v)^{3/2} g^{-1/2}$ marks the boundary of the weak coupling BCS regime.
This is dominated by the regions close to the cutoff, and yields 
\begin{eqnarray}
P^A_{SC}(g) &\sim& \exp \left( - \frac{  v^3 f(g)}{\omega_D^3 R^3} \right) ~,   \label{eq: pasc} \\
f\big(g<g_1 \big) &\sim& \frac{\Lambda^2}{\mu_0^2} \exp\big(\frac{3 \hbar^3 v^3}{g \Lambda^{2}}\big) ~, \quad f\big(g_1 < g < g_c\big) \sim \frac{ \hbar^3 v^3}{g \mu_0^2}. \nonumber
\end{eqnarray}
Here $g_1 = \frac{(\hbar v)^3}{\Lambda^{2}}$ and $g_c$ is the critical coupling for superconductivity in the clean system. 
This density of superconducting islands is doubly exponentially small in $g$ for $g \rightarrow 0$ when the islands have to be exponentially large; 
but is only exponentially small in $g$ for intermediate $g$, when small superconducting islands can form. 

In the intermediate range of $g$, the energy scale for local Cooper pairing in each island is of order $\hbar\omega_D$. However, the sample will exhibit global superconductivity only if different islands establish phase coherence. The Josephson coupling between distant islands $J$ may be determined by generalizing the calculation in \cite{Gonzalez} to the 3D Dirac point. We find that $J \sim 1 /r^5$.
Since the Josephson coupling falls off with distance faster than $1/r^3$, the coupling between nearest neighbor islands dominates. The system of locally superconducting islands embedded in a semimetal then establishes global phase coherence on temperature scales smaller than the typical nearest neighbor Josephson coupling. This leads to an estimated critical temperature for phase ordering
\begin{equation}
T^A_{c} \sim \omega_D / r^5 \sim \omega_D P_{SC}^{5/3} \sim \omega_D \exp \left( - \frac{5 v^3 f(g)}{3  R^3 \omega_D^3 } \right) ~. \label{eq: tcunbounded}
\end{equation}
Meanwhile, with model B disorder we obtain 
%
%
%
\begin{equation}
T_{c}^B \sim \exp\left( - \frac{5}{6} \frac{v^3}{ R^3 \omega_D^3} \exp(\hbar^3 v^3 / g\mu_0^2)\right) \label{eq: tcbounded}.
\end{equation}
We can understand the similarity to model A at weakest $g$ by noting that model A is then also a model of `bounded' disorder, with the bandwidth supplying the bound. Unlike model A, however, model B has no intermediate regime where $T_c$ is only exponentially small.

Thus, the dirty Weyl semimetal with attractive interactions inevitably has a superconducting ground state, but the critical temperature depends sensitively on the model of disorder. We have implicitly assumed that the pairing is $s$-wave. If the `local pairing' was not $s$-wave, then the Josephson couplings would be frustrated, and the ground state would be a `gauge glass' \cite{glass}. We leave further discussion of non-$s$-wave orders to future work, noting only that in \cite{Cho} it was determined that $\delta$-function attraction in the clean system favors $s$-wave pairing.

{\it Attractive and repulsive interactions:} We now discuss the situation when Coulomb repulsion coexists with retarded attractive interactions. We assume that the Morel-Anderson condition \cite{Anderson-Morel} is satisfied, so that local pairing on islands still occurs. However, the effective Hamiltonian for the islands must now contain not only the Josephson couplings, but also charging effects (electrostatic interactions may be neglected due to screening \cite{screening}). Thus, the effective Hamiltonian for the islands is
\begin{equation}
H = \sum_{i}  (E_c  n^2_i + V_i n_i) 
+ \sum_{\langle ij \rangle} J_{ij} \cos(\phi_i - \phi_j) ~,
\end{equation}
where $i$ and $j$ label superconducting islands, $\phi_i$ is the phase of the $i^{th}$ island, and $n_i = i \partial/\partial \phi_i.$ 
The Josephson couplings $J_{ij}$ operate primarily between nearest-neighbor islands, as previously discussed, and the $V_i n_i$ term reflects the random 
scalar potential on the islands. 
%
Such Hamiltonians have been long discussed in the theory literature \cite{Fisher, Svitsunov, Altman}, and are known to support 
a superconducting phase, and also a Bose glass \cite{Mott}. The glassy phase is characterised by an infinite superconducting susceptibility, but no long range order, 
and has a regime of stability that grows larger as the system becomes more disordered.

{\it Conclusions:} We have shown that even though weak disorder is perturbatively irrelevant at 3D Dirac points, non-perturbative effects coming from rare regions 
endow the system with a non-zero density of states and a finite mean free path. We have constructed a scaling theory of transport in the disordered semimetal, 
based on analysis of the Dirac equation in a random scalar potential. We have also considered the interplay of interactions with disorder. Purely repulsive interactions suppress rare regions, whereas purely attractive interactions lead to a ground state that is a granular superconductor, with a critical temperature that depends sensitively on the model of disorder. Meanwhile, a combination of Coulomb repulsion and retarded attractions leads to a model of Josephson coupled superconducting islands, with charging energies and random scalar potential, which can support a ground state that is a Cooper pair Bose glass. 

{\it Acknowledgements:}  We thank S.A. Parameswaran for a useful discussion.
This research was supported in part by the National Science Foundation under Grants No. DMR08-19860 (DAH) and DMR 10-06608 (SLS), and by a PCTS fellowship (RN).

\begin{widetext}

\newpage

\section{Supplementary material for `Dirty Weyl fermions'}

The $3D$ Weyl semimetal with two Dirac points has a low energy theory which is simply the massless Dirac equation. We seek insight into the behavior of the Weyl semimetal in a random scalar potential by solving the massless Dirac equation for a spherical potential well of depth $\lambda$ and radius $b$, in a geometry of radius $\rho$ (for radii $>\rho$ we will turn on a large mass, such that all states are localized to $r<\rho$). We work in units $\hbar = 1, v=1$. We solve the Dirac equation in this `spherical potential well' by separately solving the Dirac equation  in the regions $r < b$ and $r > b$, and then matching solutions at the boundary. Since we are dealing with a first order differential equation, the only matching condition is that the wave function must be continuous at the boundary.

The massless Dirac equation can be written as $(-i \hbar v \alpha_i \partial_i + V(r) ) \psi = E  \psi$, where $\psi$ is a four component spinor, $i=1,2,3$, and
\begin{equation}
\alpha_i = \left(\begin{array}{cc} 0 &  \sigma_i \\ - \sigma_i & 0 \end{array} \right).
\end{equation}
Here $\sigma_i$ are the usual Pauli matrices and we have chosen to work in the Weyl representation, and have also assumed that the Fermi velocity is isotropic. If the Fermi velocity were anisotropic, the analysis would follow through in the same way, but instead of considering a spherical potential, we would have to consider an ellipsoidal potential (such that we could rescale co-ordinates to make the potential spherically symmetric and the Fermi velocity isotropic). 

Our treatment of the 3D Dirac equation follows \cite{Callan}, but uses the Weyl representation rather than the Pauli Dirac representation. In the absence of a mass, the two Dirac points decouple. A scalar potential couples to the sum of the densities on the two Dirac points, and thus also does not couple the two Dirac points. We can thus separately solve for states near the two Dirac points. In terms of the two component spinors $\psi_{\pm \vec{Q}}(\vec{r})$ describing states near the $\pm$ Dirac points, the Dirac equation can be written as
\begin{equation}
\big(\pm i \hbar v \sigma_i \partial_i + V(\vec{r}) - E\big) \psi_{\pm \vec{Q}}(\vec{r}) = 0
\end{equation}

The eigenstates of the Hamiltonian are also eigenstates of total angular momentum $j$, but are not eigenstates of orbital angular momentum $l$. Using the standard Pauli matrix multiplication identity $\sigma_i \sigma_j = \delta_{ij} + i \epsilon_{ijk} \sigma_k$, we rewrite the gradient term as
\begin{equation}
\sigma_i \partial_i = \frac{\sigma_i r_i}{r_j r_j} \sigma_k r_k \sigma_l \partial_l = \frac{\vec{\sigma}\cdot \hat{r}}{r}(r_l \partial_l + i \epsilon_{klm} r_k \partial_l \sigma_m) = \frac{\vec{\sigma}\cdot \hat{r}}{r}\big(r \frac{\partial}{\partial r} + i \vec{\sigma} \cdot ( \vec{r} \times \vec{\partial})\big) = \vec{\sigma}\cdot \hat{r} \big(\partial_r - \frac{\vec{\sigma} \cdot \vec{L}}{\hbar r}\big)
\end{equation}
using the notation $\hat{r} = \vec{r}/r$ and $r^2 = r_j r_j$, and where $\vec{L}$ is the usual quantum mechanical angular momentum operator. This prompts us to search for a solution of the form $\psi = R(r) \phi$, where $R$ is a scalar function that depends purely on radius, whereas $\phi$ is a two component spinor which is an eigenstate of the angular momentum operator, and which is independent of radius.

Now, the eigenstates of the operator $\vec{\sigma} \cdot \vec{L}$ are two component spinors $\phi^{\pm}_{j,j_z}$ with total angular momentum $j$, angular momentum projection onto the z-axis $j_z$, and orbital angular momentum $l = j \mp 1/2$, which take the explicit form \cite{Callan}
\begin{equation}
\phi_{j, j_z}^{\pm}  = \left( \begin{array}{c} \sqrt{\frac{l+1/2 \pm j_z}{2l+1}} Y^l_{j_z-1/2} \\ \pm \sqrt{\frac{l+1/2 \mp j_z}{2l+1}} Y^l_{j_z+1/2} \end{array} \right)
\end{equation}
where the $Y$ functions are the usual spherical harmonics. We note that the $\pm$ superscript refers to the angular structure, not to which Dirac point we are on. Using the identities $\vec{J} = \vec{L} + \frac12 \vec{\sigma}$ and $\vec{J}\cdot \vec{J} = j(j+1) \hbar^2 $, $\vec{L} \cdot \vec{L} = l(l+1) \hbar^2$, we can show that the spinors obey $\vec{\sigma} \cdot \vec{L} \phi^{\pm}_{j, j_z} = - (1+ \kappa)\hbar  \phi^{\pm}_{j, j_z}$, where $\kappa = -(j+1/2)$ is a negative integer for $\phi^+$ and $\kappa = j+1/2$ is a positive integer for $\phi^-$.

We note that the functions $\phi_{j, j_z}^{\pm}$ have orbital angular momentum differing by one, and thus have opposite parity under inversion. Since $\vec{\sigma} \cdot \hat{r}$ commutes with the angular momentum operator and changes sign under inversion, it follows that it must turn $\phi^+$ into $\phi^-$ and vice versa. Since the gradient term mixes the angular sectors $\phi^{\pm}$, the eigenstates of the Hamiltonian must be linear superpositions of pieces with $\phi^+$ and $\phi^-$ angular structure. Translating \cite{Callan} to the Weyl basis, we find that the eigenstates in the vicinity of the Dirac points at $\pm \vec{Q}$ take the form
%
\begin{equation}
\psi_{+\vec{Q}} = f(r) \phi^{\pm}_{j, j_z} +i  g(r) \phi^{\mp}_{j, j_z}; \qquad \psi_{-\vec{Q}} = f(r) \phi^{\pm}_{j, j_z} -  i g(r) \phi^{\mp}_{j, j_z}
\end{equation}
where $f$ and $g$ are purely radial functions with no angular dependence.

We work with the $+\vec{Q}$ Dirac point for specificity, noting that for every state on this Dirac point there is a corresponding state on the $-\vec{Q}$ point. We note that all states at each Dirac point come in degenerate pairs that differ only in their angular structure. At the $+\vec{Q}$ Dirac point, we have states $f \phi^+ +  i g \phi^-$ and $f \phi^- + i g \phi^+$, and similarly there are two states at the $-\vec{Q}$ Dirac point. Substituting this expression for the wavefunctions into the Dirac equation (2) leads to the two equations
\begin{equation}
\frac{1}{\hbar v}(E - V) f = \partial_r g + \frac{1- \kappa}{r} g ; \qquad \qquad -\frac{1}{\hbar v}(E-V)g = \partial_r f + \frac{1+\kappa}{r} f
\end{equation}
where $\kappa$ is a positive integer for one solution, and $\kappa$ is a negative integer for its degenerate partner which differs only in its angular structure. Let us pick positive $\kappa$ for specificity. Some elementary manipulations lead to
\begin{equation}
r^2 \partial^2_r f + 2r  \partial_r f +\big( \frac{(V-E)^2 r^2}{\hbar^2 v^2} -  \kappa (1+\kappa)\big)f =0
\end{equation}
For a uniform $V$, we recognize this as the spherical Bessel equation, whose solutions are spherical Bessel functions. Substituting $f$ back into the equation for $g$ then determines $g$. Thus, the solutions for arbitrary $E \neq V$ take the form
\begin{eqnarray}
f(r) = \frac{A}{\sqrt{|V-E|r/\hbar v}}J_{\kappa+1/2}\big(|V-E|r/\hbar v\big) +  \frac{B}{\sqrt{|V-E|r/\hbar v}}K_{\kappa+1/2}\big(|V-E|r/ \hbar v\big) \\
g(r) = \sign(V-E)\left( \frac{A}{\sqrt{|V-E|r/\hbar v}} J_{\kappa - 1/2}(|V-E| r/ \hbar v) +  \frac{B}{\sqrt{|V-E|r/ \hbar v}}K_{\kappa-1/2}\big(|V-E|r/ \hbar v\big) \right)
\end{eqnarray}
where $J$ and $K$ are Bessel functions of the first and second kind respectively. We note that this is dimensionally correct, since $\hbar v$ has dimensions $Energy \times distance$, i.e. $\hbar v \approx \Lambda a$, where $\Lambda$ is the bandwidth and $a$ is the lattice scale. To save writing, we now adopt a system of units where $\hbar v = 1$. We will re-introduce $\hbar v$ whenever necessary for clarity.

 For $r<b$, $V= \lambda$. In this region, we must have $B=0$ to have a regular solution at the origin. Meanwhile, for $r>b$, $V = 0$. In this region we can have $A' \neq 0$ and $B' \neq 0$. Thus, we have
\begin{eqnarray}
f(r) = \frac{A}{\sqrt{|\lambda-E|r}}J_{\kappa+1/2}\big(|\lambda-E|r\big) \Theta(b-r) + \left(\frac{A'}{\sqrt{|E|r}}J_{\kappa+1/2}\big(|E|r\big) +  \frac{B'}{\sqrt{|E|r}}K_{\kappa+1/2}\big(|E|r\big) \right) \Theta(r-b)\nonumber \\
g(r) = \sign(\lambda-E) \frac{A}{\sqrt{|\lambda-E|r}} J_{\kappa - 1/2}(|\lambda-E| r) \Theta(b-r) - \sign(E) \left( \frac{A'}{\sqrt{|E|r}} J_{\kappa - 1/2}(|E| r) +  \frac{B'}{\sqrt{|E|r}}K_{\kappa-1/2}\big(|E|r\big) \right)\Theta(r-b)\nonumber \end{eqnarray}

Although we apparently have 3 undetermined constants $A, A'$ and $B'$, all three constants will be fixed by matching conditions at the boundary $r=b$ and by overall normalization.

Since we are dealing with a first order differential equation, only the wave function need be continuous (there is no requirement that derivatives be continuous). Imposing continuity of the wave function then implies that
\begin{equation}
\left(\begin{array}{c} A' \\ B' \end{array}\right) = A \sqrt{|E|/|\lambda-E|} \frac{1}{\Delta} \left(\begin{array}{cc} K_{\kappa - 1/2}(|E|b) & -K_{\kappa + 1/2}(|E|b) \\ - J_{\kappa - 1/2}(|E|b) & J_{\kappa+1/2}(|E|b) \end{array} \right)\left( \begin{array}{c} J_{\kappa + 1/2}(|\lambda - E|b)  \\ \sign(\frac{E}{E-\lambda}) J_{\kappa - 1/2}(|\lambda - E|b)\end{array}\right)
\end{equation}
where $\Delta$ is the determinant of the $2\times2$ matrix. This fails for special values of $E$ where the matrix is singular (vanishing determinant).

It is simpler to note that continuity of the wave function also implies continuity of the probability density (given by the norm squared of the wave function). The norm squared of the wave function at $r=b$ (defined as $|f|^2 + |g|^2$) never vanishes, and scales as $(\lambda - E)^{-2} b^{-2}$ in the limit of large $|\lambda - E|b$ while saturating to a constant in the limit of small $|\lambda - E|b$. 
Thus, the probability density just outside the well never vanishes,
%
%
and there is always `leakage' of the probability density out of the region $r<b$. Moreover, the spherical Bessel functions only decay as $1/r$ at long distances, so the probability density only decays as $1/r^2$ at long distances. Thus, for general $E$, the wave function is not localized on the potential, but rather is spread through space, with most of the probability outside.

\section{Bound States from Special Wells}
Now lets consider the special case $E=0$. Now the two equations in (6) decouple, and can be straightforwardly solved to give an exterior solution
\begin{equation}
f(r>b) \sim r^{-(1+\kappa)} \textrm{ or } f(r) = 0; \qquad \qquad g(r>b) \sim r^{\kappa - 1} \textrm{ or } g(r) = 0
\end{equation}
Recall that $\kappa$ is a positive integer. This corresponds to a bound state if and only if we pick the solution $g(r) = 0$, which comes about if $g(r)$ is matched to a node of the interior Bessel function. This in turn happens only for special values of the well depth $\lambda_c$. Although there is a well depth corresponding to a bound state for all values of $\kappa$, larger values of $\kappa$ require a deeper (or wider) well in order to have a bound state, and are exponentially rarer. We therefore restrict our attention to $\kappa = 1$, which has bound state solutions $g(r) = 0$ for $\lambda = \lambda^n_c \approx n \pi / b$, where $n$ is a positive integer. Again, values of $n$ greater than one are exponentially rarer than $n=1$, so the most common bound state involves $\kappa = 1$ and $n=1$, with a well depth $\lambda_c \approx \pi / b$. The probability density in this bound state decays like $1/r^4$ outside the well (i.e. most of the probability density is localized on the well).

We note that the angular eigenfunction $\phi_-$ has total angular momentum $j = l-1/2=1/2$ (for $\kappa = 1$), but may have $j^z = \pm j$. Thus there are two bound states corresponding to the $\kappa = 1$ solution identified above. We note that there are two additional bound states corresponding to $\kappa = -1$ and $\lambda = \lambda_c = \pi/b$, which now corresponds to an angular eigenfunction $\phi_+$ and has $f(r) = 0$. Thus, there are four bound states per Dirac point for each special well. Thus, with $N$ Dirac points there are $4N$ bound states per for each special well.


In an infinite sample, even an infinitesimal deviation from $\lambda = \lambda_c$ or $E =0$ leads to most of the wave function leaking out of the well. This just tells us that a single potential well cannot generate a finite density of states in an infinite system. However, in a system with many potential wells, there will be some finite window $\delta \lambda, \delta E$, which allows for bound states. 
To make further progress requires an additional calculation.

\subsection{Scattering off a spherical potential well}
We now determine the scattering cross section $\sigma(\lambda, E)$ of the spherical potential well. This is determined as $\sigma = \frac{4\pi}{k^2} \sin^2 \delta = \frac{4 \pi \hbar^2 v^2 }{E^2} \sin^2 \delta$, where $\delta$ is the phase shift \cite{Perkins}. In the absence of a scattering potential, the solution would be purely a spherical Bessel function of the first kind, which at long distances has the asymptotic form $J_{\alpha}(kr) \sim \frac{1}{\sqrt{kr}} \cos(kr - \alpha \pi/2 - \pi/4)$. In the presence of a scattering potential, the solution is $\psi \sim \big(A' J_{\alpha}(kr) + B' K_{\alpha} (kr) \big) \sim A' \cos(kr - \alpha \pi/2 - \pi/4) + B' \sin(kr - \alpha \pi/2 - \pi/4) \sim C \cos (kr - \alpha \pi/2 - \pi/4 - \delta)$, where $\delta$ is the phase shift. Application of standard trigonometric identities, as well as Eq.(10), then leads to \cite{Dombey}

\begin{equation}
\tan \delta = \frac{\sign \big(\frac{E}{E-v}\big)J_{3/2}(|E|b) J_{1/2}(|E-V| b) - J_{1/2}(|E| b) J_{3/2}(|E-V| b )}{\sign \big(\frac{E}{E-V}\big) J_{1/2}(|E-V| b) K_{3/2}(|E| b) - J_{3/2}(|E-V| b) K_{1/2} (|E| b)}
\end{equation}
This has resonances at critical values of the well depth, corresponding to phase shifts $\delta = \pi/2$. Substituting into the expression for the cross section, we find that in the scaling limit $E \rightarrow 0$, the cross section is a tightly peaked Lorentzian, with
\begin{equation}
\sigma(E, \lambda) \sim \frac{\hbar^2 v^2 E^2 b^2 }{(\lambda - \lambda_c(E))^2\hbar^2 v^2  + E^4 b^2}; \qquad \lambda_c(E) - \lambda_c(0) \sim E; \qquad \lambda_c(0) \sim \pm \pi / b
\end{equation}
where, we recall, $\hbar v = \Lambda a$. From this we conclude that there is a line of resonances in the $\lambda, E$ plane, and that these resonances have width $\sim b E^2$ in both $\lambda$ and $E$. This leads us to the $\delta \lambda \sim \delta E \sim E^2$ scaling we presented in the main text.

\subsection{Scaling theory}

We can now revisit the density of special wells. At an energy $E$, there is a density $P(\lambda_c) \delta \lambda = \nu_0 \delta \lambda \sim \nu_0 b E^2$ of special wells
that are near-resonant at that energy, and these resonances are spread over a bandwidth $\delta E \sim b E^2$. Thus the density of states is $\nu_0 \delta \lambda / \delta E = \nu_0$, which is independent of $E$. We have not taken the non-resonant `extended' states into account, thus this estimate is valid only on scales $E < \sqrt{\nu_0}$, where the density of states $\nu_0$ from special wells exceeds the DOS $\sim E^2$ from the extended states.

Meanwhile, the mean free path from scattering off resonant wells behaves as
\begin{equation}
l \approx \frac{1}{\int d \lambda P(\lambda) \sigma(\lambda, E)} \sim (\nu_0 b) ^{-1} ~.
\end{equation}
Note that for Model A disorder, $b \sim R$, where $R$ is the correlation length for the disorder, whereas for Model B disorder, $b \sim 1/\mu_0$, where we have used the terminology introduced in the main text.
We are considering weak disorder, where $V=\mu_0^2 b^3 <b$ for either model of disorder.

Applying the Ioffe-Regel criterion \cite{IoffeRegel} $k l = E l = 1$, we find that for $E < \nu_0 b $, we are in the `strong scattering' regime where it no longer
makes sense to talk about weakly-scattered extended states.  In this regime, the states all live on rare regions, and transport proceeds by hopping.
In this region we have $\delta \lambda \sim \delta E \sim b^3 \nu_0^2$, and the typical hopping rate is also $b^3 \nu_0^2$.
Meanwhile, the density of rare regions is $P(\lambda_c) \delta \lambda \sim b^3 \nu_0^3$, and the typical spacing is $(b \nu_0)^{-1}$.
Thus, transport in this regime occurs due to hopping over length scales $(b \nu_0)^{-1}$, consistent with the result quoted in the main text. 

Meanwhile, at high energy where the SCBA remains valid, the resulting mean free path is $l\sim 1/(VE^2)$.  The
rare regions start to dominate the scattering when this SCBA mean free path exceeds that due to the rare regions, which is at an energy scale $ E \lesssim \sqrt{\nu_0 b/V}$, but they do not start to dominate the density of states until $E \lesssim \nu_0^{1/2}$.  Moreover, we do not enter the strong scattering / hopping conduction regime until $E < \nu_0 b$ (according to the Ioffe-Regel criterion). Thus, we are led to identify two distinct intermediate energy regimes. In the regime $\nu_0 b < E < \nu_0^{1/2}$, the DOS and scattering are dominated by the rare regions, but the mean free path is still much longer than the wavelength and the scattering is in this sense weak. Meanwhile, for $\nu_0^{1/2} < E < (\nu_0 b/V)^{1/2}$, the DOS is dominated by extended states, but the (still weak) scattering is dominated by rare regions.

In both intermediate energy regimes, 
the carriers spend a typical time $\sim b^{-1} E^{-2}$ trapped on each resonant special well (this is just the width of the resonance) whereas the time spent traveling
freely in between special wells is proportional to the mean free path $l \sim (b\nu_0)^{-1}$. 
Thus, in the first intermediate energy regime $\nu_0 b < E < \nu_0^{1/2}$, the time spent trapped on resonances is much longer than the time spent traveling freely,
whereas in the second intermediate energy regime $\sqrt{\nu_0} < E < \sqrt{\nu_0 b/V}$, the time spent traveling freely exceeds the time spent trapped on resonances.


Some properties of each of our four regimes are summarized in Table I.  In each case the diffusivity is $D\sim l^2/\tau$, with $l$ the typical hopping distance in the hopping regime and the mean free path in the other regimes.
The time between hops or scattering events is $\tau$.  The  zero-temperature conductivity for these noninteracting carriers is then
$\sigma_{DC} =  \nu e^2 D$, where $\nu$ is the DOS. 
Stitching together the low energy (hopping dominated) and high energy (SCBA) regions leads to the plot Fig.1. 

\begin{table}
\begin{tabular}{|c|c|c|c|c|c|c|}
\hline
Energy regime & Description & Length scale & Time scale & DOS & Diffusivity & DC conductivity\tabularnewline
\hline
$E <  (\hbar v)^2 \nu_0 b$ & Hopping & $(\hbar v \nu_0 b)^{-1}$ &  $(\hbar^2 v^3 \nu_0^2 b^3)^{-1}$& $ N \nu_0 $& $v b $ & $N e^2 \nu_0 v b$\tabularnewline
\hline
$(\hbar v)^2 \nu_0 b < E < (\hbar v)^{3/2} \nu_0^{1/2} $ & Intermediate I  & $(\hbar v \nu_0 b)^{-1}$ & $\frac{\hbar^2 v}{E^2 b}$ & $ N \nu_0 $& $\frac{E^2}{\hbar^4 v^3 b \nu_0^2 } $ & $\frac{Ne^2  E^2}{ \hbar^4 v^3 b \nu_0 } $\tabularnewline
\hline
$(\hbar v)^{3/2} \nu_0^{1/2} < E < (\hbar v)^{5/2} ( \nu_0 b/V)^{1/2} $ & Intermediate II  & $(\hbar v \nu_0 b)^{-1}$ & $(\hbar v^2 \nu_0 b)^{-1}$ & $ N \frac{E^2}{(\hbar v)^3} $& $\frac{1}{\hbar b \nu_0}$ & $\frac{Ne^2 E^2}{ \hbar^4 v^3 b \nu_0} $\tabularnewline
\hline
$(\hbar v)^{5/2} ( \nu_0 b/V)^{1/2} < E$ & SCBA & $\frac{(\hbar v)^4}{V E^2}$ & $\frac{\hbar^4 v^3}{V E^2}$ & $ N \frac{E^2 }{(\hbar v)^3}$& $\frac{\hbar^4 v^5}{V E^2} $ & $N \frac{e^2}{\hbar} \frac{(\hbar v)^2}{V}$ \tabularnewline
\hline
\end{tabular}
\caption{Table listing the scaling properties of the four distinct energy regimes (up to purely numerical prefactors). Here $N$ is the number of Dirac points,
$v$ is the Fermi velocity, and $\nu_0$ is the (exponentially small) zero energy density of states per unit volume, calculated in the main text.
We have used $V = \mu_0^2 R^3$ to denote the mean square scalar potential, where $R$ is the correlation length of the disorder
(which is of order the lattice constant). Finally, $b$ is the typical radius of a special well. For Model A disorder, $b \sim R$, whereas for model B disorder,
$b \sim \hbar v/\mu_0$. In the limit of small $\mu_0$ (i.e. weak disorder), $\hbar^2 v^2 b/V \gg 1$.  The `Length scale' column lists the typical hopping distance
in the hopping regime, and the mean free path in all other regimes. The `Time scale' column lists the typical hopping time in the hopping regime, the
typical dwell time on a resonant well in intermediate regime I, and the scattering time in the other two regimes.
The rest of the columns seem self explanatory.  
For the estimates of the transport in the hopping regime, we assume those states are not localized and the carriers do a random walk with the step length and time set
by these scales; this is what happens in the other regimes.}
\end{table}

We note that when estimating the diffusion constant using the method of continuous time random walks, we ignore the possibility of destructive interference between distinct paths. Such destructive interference could give rise to localization in the hopping model at very long lengthscales (very low energy scales). We do not investigate this issue further here, leaving it as a topic for future work. We note however that in the hopping regime, the conductance at the length scale of a typical hop is $N e^2/\hbar$, where $N$ is the number of Dirac points, and thus the system may be close to an Anderson transition. The possibility of localization in the hopping model at the lowest energies is thus a fruitful topic for future work. The localization (or lack thereof) may also be sensitive to the symmetries of the problem, and thus may discriminate between Weyl semimetals and other more general 3D Dirac points, unlike the other features investigated in this work.

\end{widetext}


\begin{thebibliography}{99}
\bibitem[internode (2013)]{internode}
We focus on random scalar potential disorder, which produces intranode scattering. We neglect internode scattering effects.
\bibitem[bounded (2013)]{bounded}
This heuristic argument gives $\nu \sim R^{5/2} \exp{(-1/VR)}$. In the white noise limit $R\rightarrow 0$, the resonant regions are not exponentially rare, but their contribution to the DOS vanishes as $R^{5/2}$, because of the pre-factor. However, it is not clear whether the pre factor is accurately estimated by such an argument, thus the exact behavior in the white noise limit remains an open question. 
\bibitem[Mott (2013)]{Mott}
With bounded disorder in $V_i$, the large $E_c$ limit may also support a phase with `Mott insulating' islands embedded in the semimetal.
\end{thebibliography}
\end{document}